\documentclass[conference,a4paper]{IEEEtran}
\IEEEoverridecommandlockouts
\usepackage{cite}
\usepackage{amsmath,amssymb,amsfonts}
\usepackage{algorithmic}
\usepackage{graphicx}
\usepackage{textcomp}
\usepackage{xcolor}
\usepackage{color,soul}
\usepackage{graphicx, caption, subcaption}
\def\BibTeX{{\rm B\kern-.05em{\sc i\kern-.025em b}\kern-.08em
    T\kern-.1667em\lower.7ex\hbox{E}\kern-.125emX}}

\usepackage{comment}

\begin{document}

\title{Cryo-CMOS Antenna for Wireless Communications within a Quantum Computer Cryostat

\thanks{Authors gratefully acknowledge funding from the European Commission via projects with GA 101042080 (WINC), 101099697 (QUADRATURE). }
}

\author{\IEEEauthorblockN{
Viviana Centritto\IEEEauthorrefmark{1},   
Ama Bandara\IEEEauthorrefmark{1},   
Heqi Deng\IEEEauthorrefmark{2},    
Masoud Babaie\IEEEauthorrefmark{2},      
Evgenii Vinogradov\IEEEauthorrefmark{1},      
Sergi Abadal\IEEEauthorrefmark{1}, \\      
Eduard Alarc\'on\IEEEauthorrefmark{1}    
}                                     
\IEEEauthorblockA{\IEEEauthorrefmark{1}NaNoNetworking Center in Catalunya, Universitat Polit\`{e}cnica de Catalunya Barcelona, Spain}
\IEEEauthorblockA{\IEEEauthorrefmark{2} 
  Department of Microelectronics and the Department
of Quantum and Computer Engineering, Delft University of Technology, \\ CD Delft, The Netherlands}
 \IEEEauthorblockA{ \emph{*ama.peramuna@upc.edu} }
}

\maketitle

\begin{abstract}
Scaling quantum computers from a few qubits to large numbers remains a challenge in realizing practical quantum advantage. Multi-core quantum architectures offer a promising solution by enabling scalability through distributed quantum processing units (QPUs) interconnected via classical and quantum links. However, the bottleneck of interconnects persists, as dense wiring across temperature stages and within the same layer introduces spatial constraints, power dissipation, and latency. To address this, we propose a cryo-compatible on-chip differential dipole antenna operating at 28 GHz to enable short-range wireless communication within a quantum computer cryostat. Temperature dependent material properties are incorporated to model antenna behavior at 4 K. Furthermore, by embedding the antenna into a realistic cryostat structure, we assess its feasibility within the cryogenic environment. The proposed antenna achieves a reflection coefficient of -28 dB in free space and -21 dB within the cryostat, demonstrating efficient impedance matching.
\end{abstract}

\begin{IEEEkeywords}
cryogenic on-chip antenna, wireless interconnects, cryostat environment, multi-core quantum computing
\end{IEEEkeywords}

\section{Introduction}

Quantum computing is a revolutionary technology that uses the unique properties of qubits, namely superposition and entanglement, to tackle complex problems far beyond the reach of today’s classical computers. However, to unlock its full potential and achieve the quantum advantage in solving real-world challenges, quantum systems must scale from thousands to millions of qubits, thereby surpassing the capabilities of classical supercomputers \cite{Ladd2010}.

The scalability of quantum computers captures a variety of aspects considering the footprint of the system at large, managing the scaled-up classical-quantum interconnects, cross-talk, and power dissipation. 
The classical interface provides the capability of exchanging information from a large number of qubits through wired connections with the classical computer at room temperature, for control and read-out, with input-output electrical signals. Therefore, with each wire connected to a single qubit in a densely packed Quantum Processor Unit (QPU), the proportional increment of interconnects will result in thermal, power, and spatial constraints. Even though wired interconnects would provide high speed data transfer with low energy rates, exhausting the delicate cryogenic environment with densely packed wires would also increase the thermal profile, which leads to qubit de-coherence.

As an alternative to the scaled-up monolithic quantum computing system, recently modular quantum architectures emerge as a profound solution \cite{Brown2016, Isailovic2006,Monroe2014,Alarcon2023}. In them, several QPUs are utilized for the purpose of parallel processing of the quantum algorithm, by connecting through both quantum and classical links. However, in such system the interconnect fabric plays a crucial role in orchestrating the top-down vertical connections though each temperature stage and communicating in between the cores through quantum coherent links while control and synchronizations are done by the classical communications \cite{escofet2023}.

Nevertheless, the recent advancement of fabricating cryo-CMOS electronics has paved the way of integrating the control electronics at 4~K, with ultra-low power dissipation \cite{charbon2021, patra2018}. With these advancements, RF communication inside the quantum structure makes a compelling case for multi-core quantum systems, with the necessity of global connectivity and reduced latency despite of the distance, while proving reconfigurable links with broadcasting capabilities. In addition, with the integrated RF on-chip antenna and transceiver circuits at 4~K, the thermal noise added in cryogenic temperature will be low, while increasing the Signal to Noise Ratio (SNR) of the wireless link. 

In this context, in recent literature, antennas operating at cryogenic environments have been gaining interest. Bouis and Febvre \cite {Bouis2008} investigated bow-tie antennas for short-distance communication in cryogenic temperature, achieving a transmission loss of less than 3 dB over a 1.5 GHz bandwidth in the 8–12 GHz range, with crosstalk below -20 dB in array configurations. A cryogenic-wireless terahertz interconnect is proposed in \cite{Wang2025}, where a horn antenna design is explored at the 60~K stage of the cryostat to obtain a large radiation aperture and direct the radiation to the on-chip patch antennas operating at 260~GHz. The antenna radiation efficiency has improved from 38\% at 300~K to 67\%-97\% given the considerable changes in copper electrical conductivity at 4~K. As cryogenic applications grow, these antennas could leverage integration techniques from on-chip designs to further enhance efficiency and scalability in extreme environments.


Therefore, in view of the constraints and merits outlined above, in this work, by considering the noise immunity on electromagnetic (EM) coupling through a differential feed to reduce the common-mode currents and cryo-CMOS transceiver RF compatibility, we propose a differential dipole antenna operating at cryogenic temperature, which will be integrated with a cryo-CMOS transmitter designed at 4~K operating at 28 GHz \cite{charbon2021, patra2018}. The main contributions of the paper are summarized as follows:

\begin{itemize}
    \item We propose an on-chip differential dipole antenna which works in cryogenic temperature with design and simulations, by evaluating the material properties compatible in cryogenic environment. 
    \item We assess the proposed antenna at 4~K within a proposed cryostat design to minimize the impact of RF interference on the nearby QPUs.
\end{itemize}

\section{Antenna Design and Integration} 
\label{sec:model}

In this section, we discuss the design of the differential dipole antenna and integration of the proposed design on cryo-CMOS technology \cite{charbon2021,patra2018}. This topology was chosen for its compatibility with differential circuits, which are preferred in IC designs due to their noise immunity \cite{cheema2013}. Assuming spin qubits operating at 20 GHz \cite{Philips2022}, and that the required spectral bandwidth gap for noise suppression is 8~GHz (to achieve a qubit infidelity of $10^{-6}$), the operating frequency of the antenna was obtained as 28~GHz \cite{Philips2022}. For the purpose of simulation as an on chip-antenna, the chip package was modeled along with the antenna design. \par


\subsection{Differential Dipole Antenna }

\begin{figure*}[t]
\centering
\begin{subfigure}[t]{0.28\textwidth}
\includegraphics[width=1.05\textwidth]{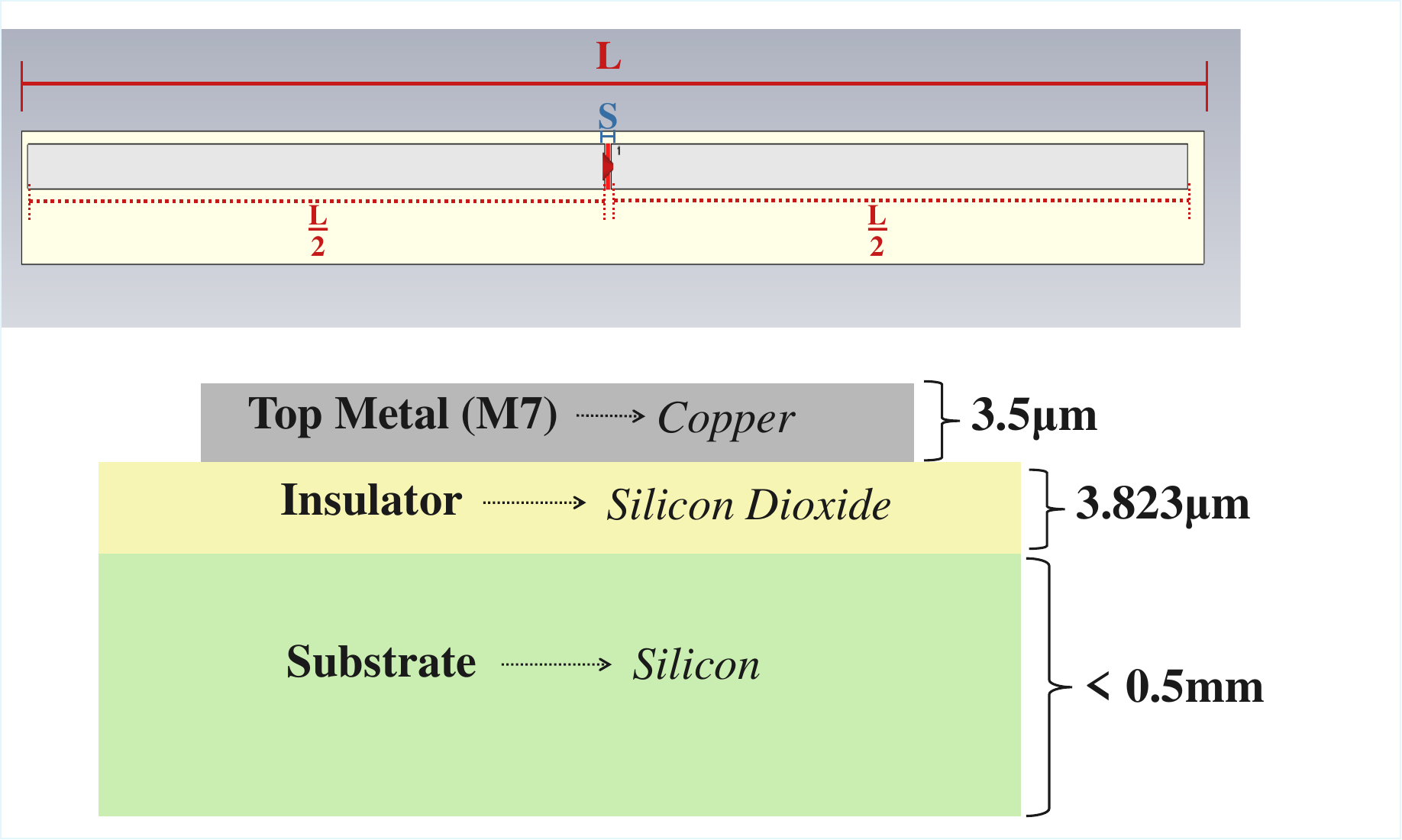}
\caption{}
\label{fig:dipoledim}
\end{subfigure}
\begin{subfigure}[t]{0.24\textwidth}
\includegraphics[width=\textwidth]{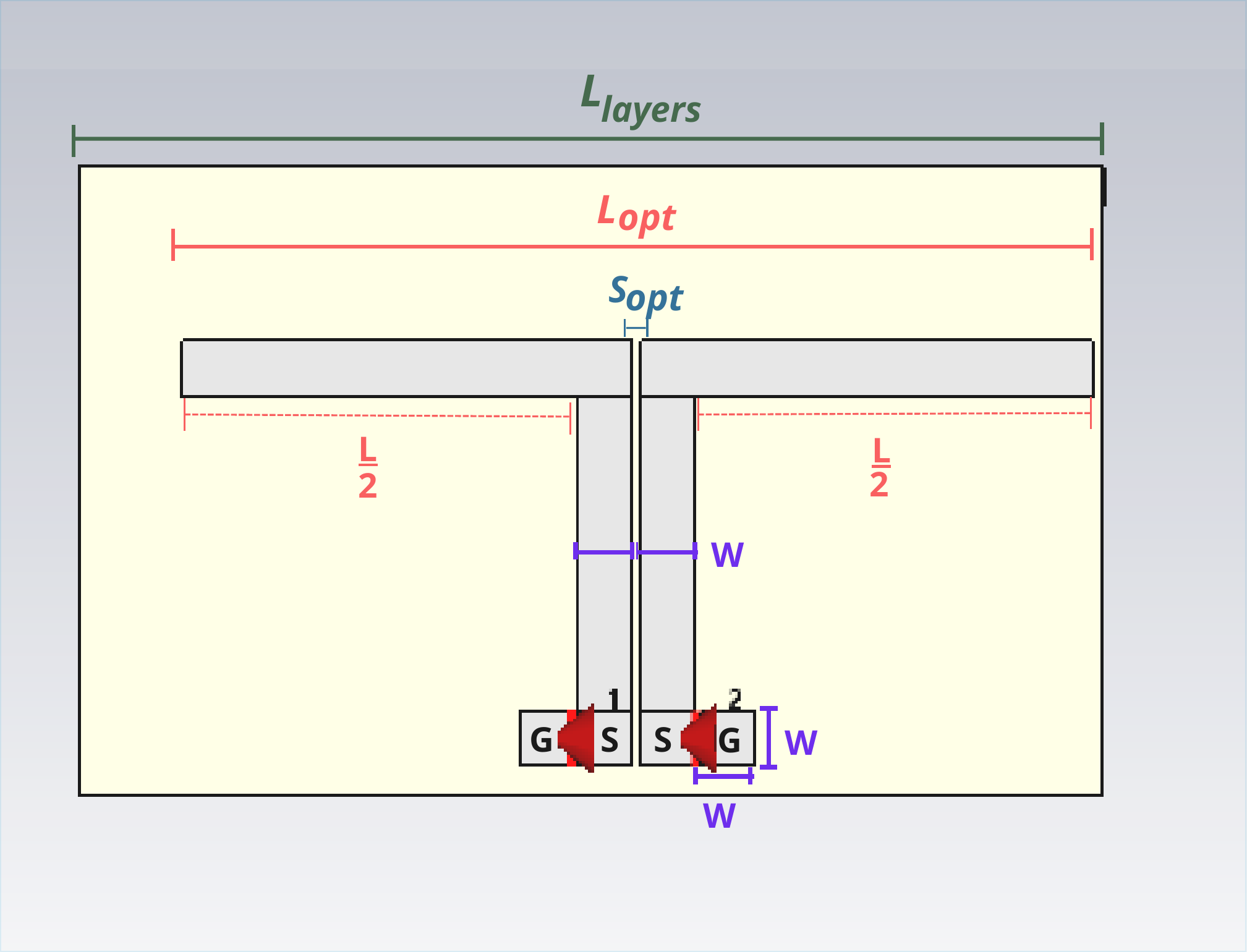}
\caption{}
\label{diffeed}
\end{subfigure}
\begin{subfigure}[t]{0.3\textwidth}
\includegraphics[width=\textwidth]{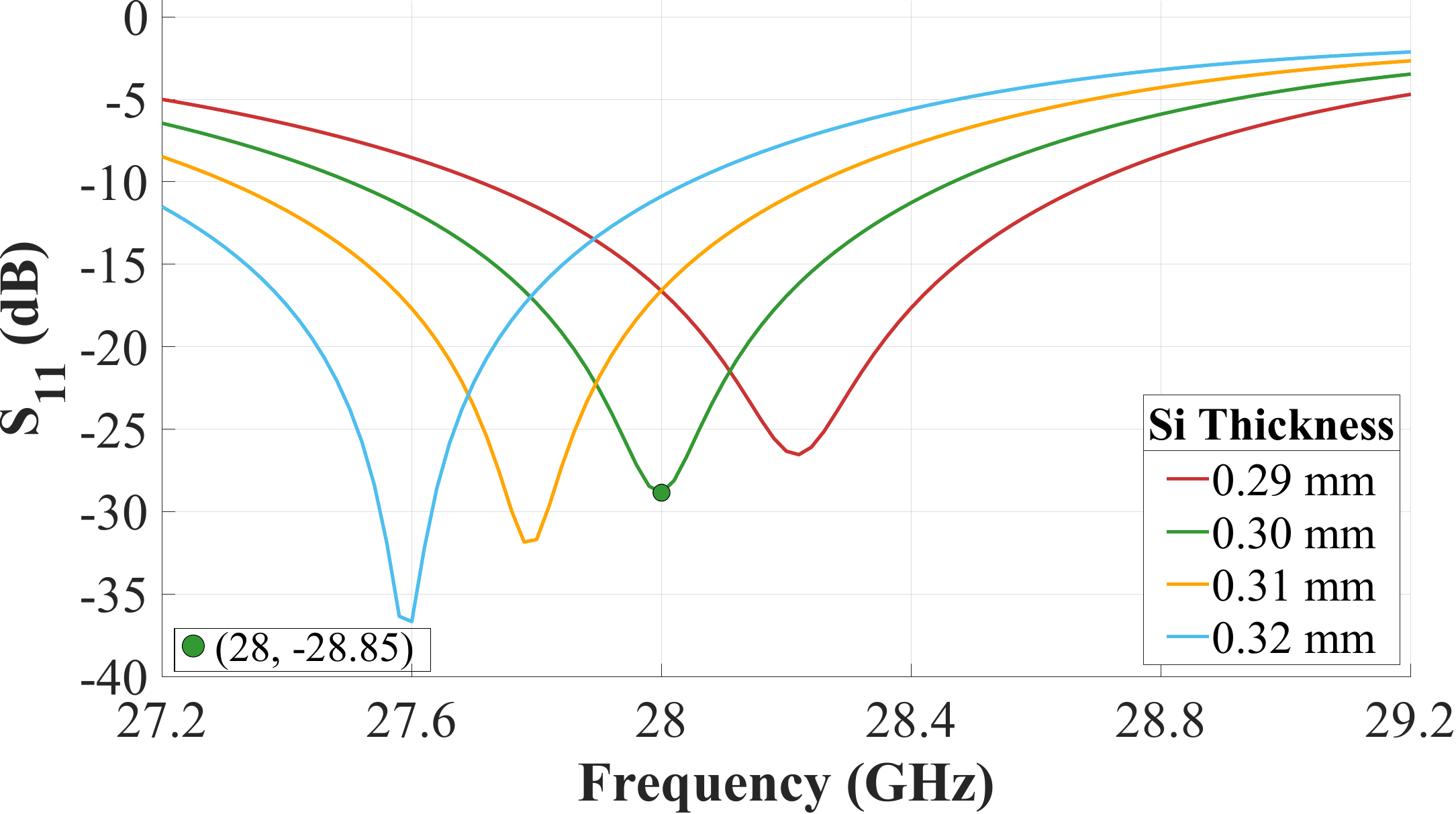}
\caption{}
\label{fig:Sithickness}
\end{subfigure}
\vspace{-0.2cm}
\caption{On-chip dipole antenna design (a) Center fed dipole design and CMOS layered stack (top) and thickness of the top metal, insulator and substrate layer (bottom) (b) Differential fed dipole design (c) Assessment of the substrate (Si) thickness.} 
\label{fig:dipoledesign}
\vspace{-0.4cm}
\end{figure*}

Fig.\ref{fig:dipoledesign} shows the schematic of the proposed antenna design. The differential dipole antenna consists of two conductive traces made of copper deposited onto the Silicon Dioxide (SiO\textsubscript{\textrm{2}}) layer. These traces serve as the two arms of the dipole, both with equal length and aligned in straight line, resembling the structure of a traditional half-wave dipole antenna. Initially a center-fed dipole was designed as shown in Fig. \ref{fig:dipoledim}, where the primary dimension of interest is the resonance length $L$, which can be theoretically obtained as
\begin{equation}
L=\frac{\lambda_{0}}{2\sqrt{\varepsilon_{eff}}},
\end{equation}
where $\lambda_{0}$ represents the free-space wavelength, and $\varepsilon_{eff}=(\varepsilon_{r}+1)/2$ corresponds to the effective relative permittivity of the material and $\varepsilon_{r}$ represents the permittivity of the substrate. 
The final dimensions were obtained by gradually adjusting the length of the dipole's arms and the distance between them.

The differential dipole feed was designed with two micro-strip lines (MLSs), where the end of each line is connected with two Ground-Signal-Signal-Ground (GSSG) square pads as shown in Fig. \ref{diffeed}. The MLSs consist of two co-planner strips located on top of the SiO\textsubscript{\textrm{2}} layer. These form a differential pair, with both strips carrying signals of the same amplitude but opposite phase. To ensure an optimal power transfer, the width of each line ($W$) and spacing between them ($S$) were calculated based on the properties and thickness of the substrate [9], where fine tuned dimensions were obtained with post-simulations.\par 

In the case of the GSSG pad configuration, the two central pads carry the differential signals, and the outer pads provide the ground reference. They are square in shape, with sides that are the same length as $W$. This layout ensures that the on-chip dipole can be measured as a stand-alone structure, isolating it from the rest of the chip's circuitry if required [10].

\subsection{Integration of the antenna On-Chip }\label{sec:chip}

The on-chip antenna consists of a layered stack, as shown in Fig. \ref{fig:dipoledim}. The dipole is located on the M7 layer, which has a thickness of $3.5~\mu$m. This dimension is consistent with the $3-4~\mu$m top metal layer typically found in CMOS stacks \cite{cheema2013}. Below that is the SiO\textsubscript{\textrm{2}} layer with a thickness of $3.823~\mu$m, which provides electrical isolation between the antenna and the substrate \cite{cheema2013}. The Silicon (Si) substrate layer is located at the bottom with an initial thickness of less than 0.5 mm \cite{cheema2013}. Modeling this last parameter is critical to the design of on-chip antennas because it affects radiation efficiency and gain.

As the on-chip dipoles are designed to be operated at cryogenic temperature, the material properties depend upon the temperature variation. We adopt the superconductive properties of the materials at cryogenic temperature to resemble the environment inside the cryostat. Table \ref{tab:material} provides the detailed description of the materials used and the change of the properties between room and cryogenic temperature \cite{patra2020,krupka2006}.

\begin{table}[t]
\centering
\caption{Material parameters}
\label{tab:material}
\begin{tabular}{|c|c|c|}
\hline
\textbf{Material Parameter} & \textbf{Cryogenic Temp. } & \textbf{Room Temp. } \\
\hline
 $Cu$ Conductivity & $2.9 \times 10^{8}$~S/m \cite{patra2020} & $5.9 \times 10^{7}$~S/m   \\
 \hline
 $Si$ Conductivity& $4.26 \times 10^{-7}$~S/m \cite{krupka2006} & $4.26 \times 10^{-4}$~S/m \\
 \hline
 $Si$ Permittivity ($\varepsilon_{r}$)& 11.45 \cite{krupka2006} & 11.75   \\
 \hline
 $SiO_{2}$ permittivity ($\varepsilon_{r}$)&3.9& 3.9  \\
 \hline
\end{tabular}%
\end{table}

\subsection{Environment Description}\label{sec:env}



The physical operations of QPUs are carried out at cryogenic temperatures, housed within a cylindrical enclosure known as a cryostat. Inside the cryostat, the required low temperatures are maintained to ensure high-fidelity qubit operations, since physically the qubits are highly sensitive to temperature fluctuations.

The cryostat consists of several temperature stages, each hosting different hardware components, with successive layers progressively cooling down to the milli-kelvin range ($\approx$4–20 mK), which is essential for stable qubit interactions. At the warmer, top-most stage reside the classical control electronics, which generate and process the control and readout signals for the qubits. As discussed in the description of dilution refrigerators \cite{Zu2022}, from top to bottom, in general the cryostat is consisting of continuous cooling, to reach the lowest temperature down to several mK required for qubit operation.

The layers of cryostat walls are carefully engineered to suppress external magnetic and thermal interference from the surrounding environment. For this purpose, thermal shielding is implemented using specific thermal insulation materials, while an additional layer of Mu-metal is applied outside the thermal shield to mitigate magnetic interference.

\begin{figure}[h]
\centering
\includegraphics[width=0.4\textwidth]{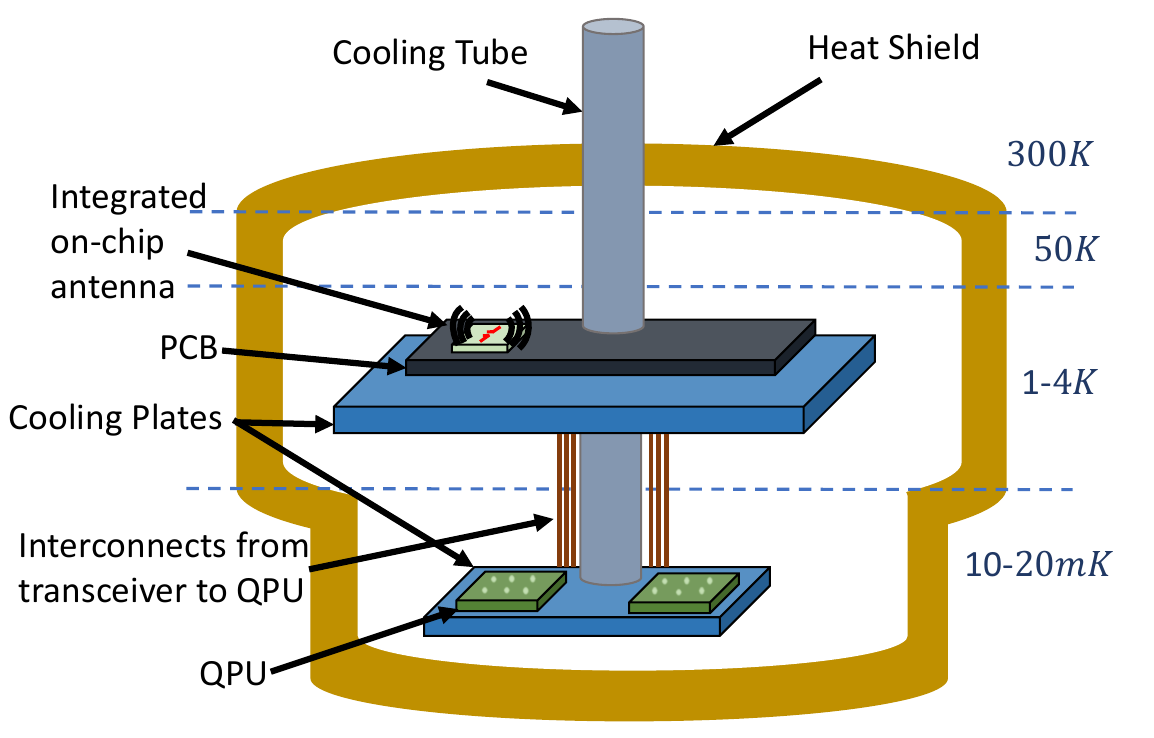}
\vspace{-0.2cm}
\caption{ The cryostat system simulation model with integrated differential dipole on-chip antennas. } 
\label{fig:model}
\vspace{-0.4cm}
\end{figure}

Fig. \ref{fig:model} illustrates the structure of the simulation model with integrated on-chip antennas. As described in Section \ref{sec:chip}, the designed differential dipole antenna for cryogenic temperatures was integrated into the cryostat structure by placing it on the printed circuit board (PCB), which is positioned above the cooling plate that is connected to the central cooling tube. The dimensions of the cryostat are based on a standard design used for quantum computing experiments in the ranges of, diameter d=30~cm and height, h=70~cm \cite{bluefors_quantum_2025}. The distance between the bottom-middle and middle-top cooling plates are 10~cm and 15~cm, respectively. 

\section{Performance Evaluation }
\label{sec:results}

To assess the performance of the proposed antenna design, we simulate the on-chip integrated differential dipole with CST MWS \cite{CST} in two scenarios, namely (i) free-space at both cryogenic and non-cryogenic temperatures, and (ii) within the cryostat. Two discrete ports on each GSSG pad were used for the excitation with impulse signals. 

\subsection{Free space}

The performance of the antenna was initially assessed in relation to the thickness of the substrate. As shown in Fig. \ref{fig:Sithickness}, the $S_{11}$ parameters are depicted across a 27-29 GHz frequency range. The reflection coefficient is -28.85 dB at 28 GHz, which corresponds to a Si thickness of  0.30 mm. This finding is compatible with the design of the cryo-CMOS electronics on the chip \cite{charbon2021,patra2018}. The dipole optimal length ($L_{opt}$) and the optimal gap between the dipoles ($S_{opt}$) are obtained as 2.8 mm and 0.03 mm respectively. 



\begin{figure*}[!t]
\centering
\begin{subfigure}[t]{0.31\textwidth} 
\includegraphics[width=1\textwidth]{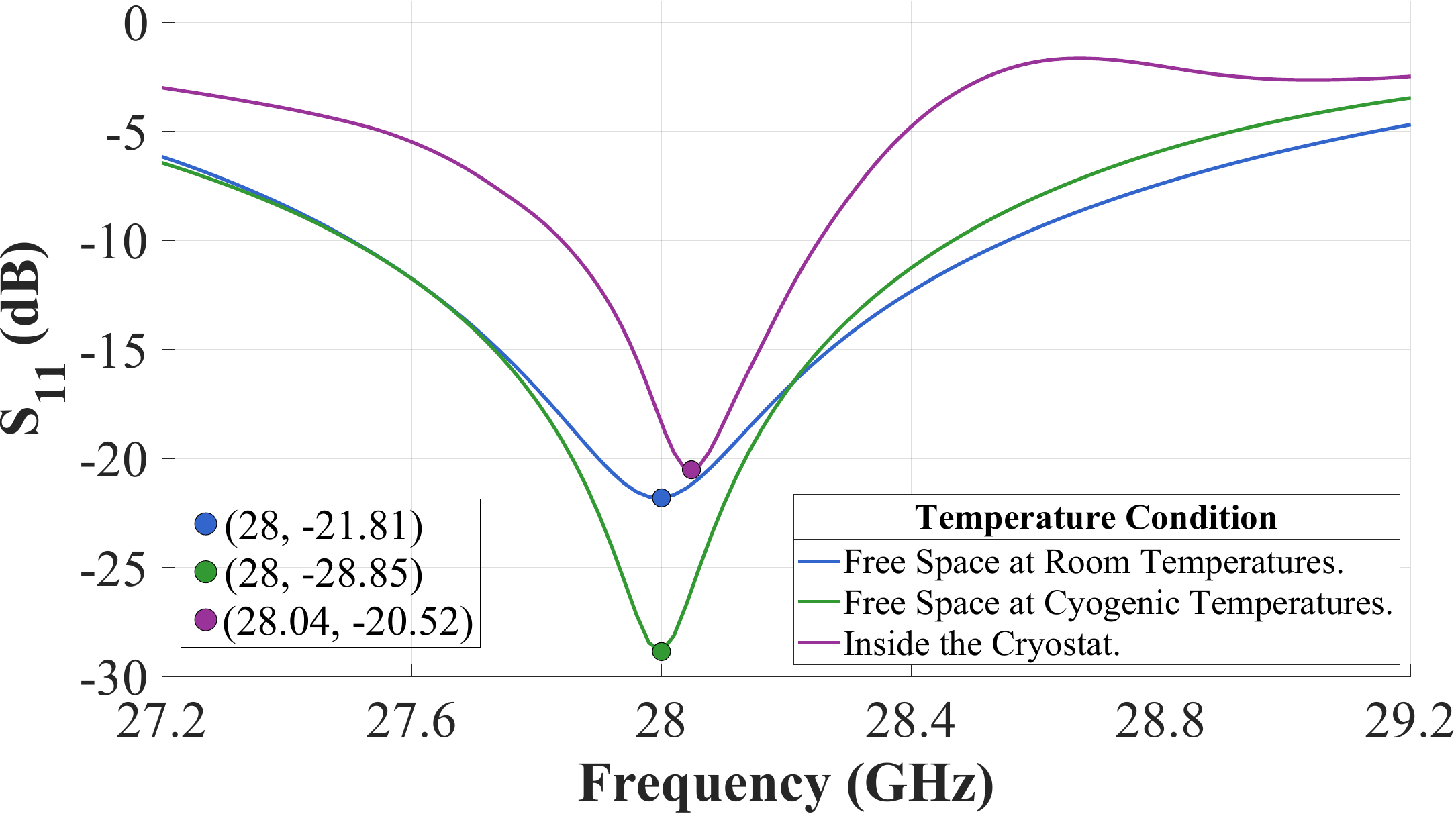}
\caption{}
\label{fig:roomvscryo}
\end{subfigure}
\begin{subfigure}[t]{0.31\textwidth} 
\includegraphics[width=1\textwidth]{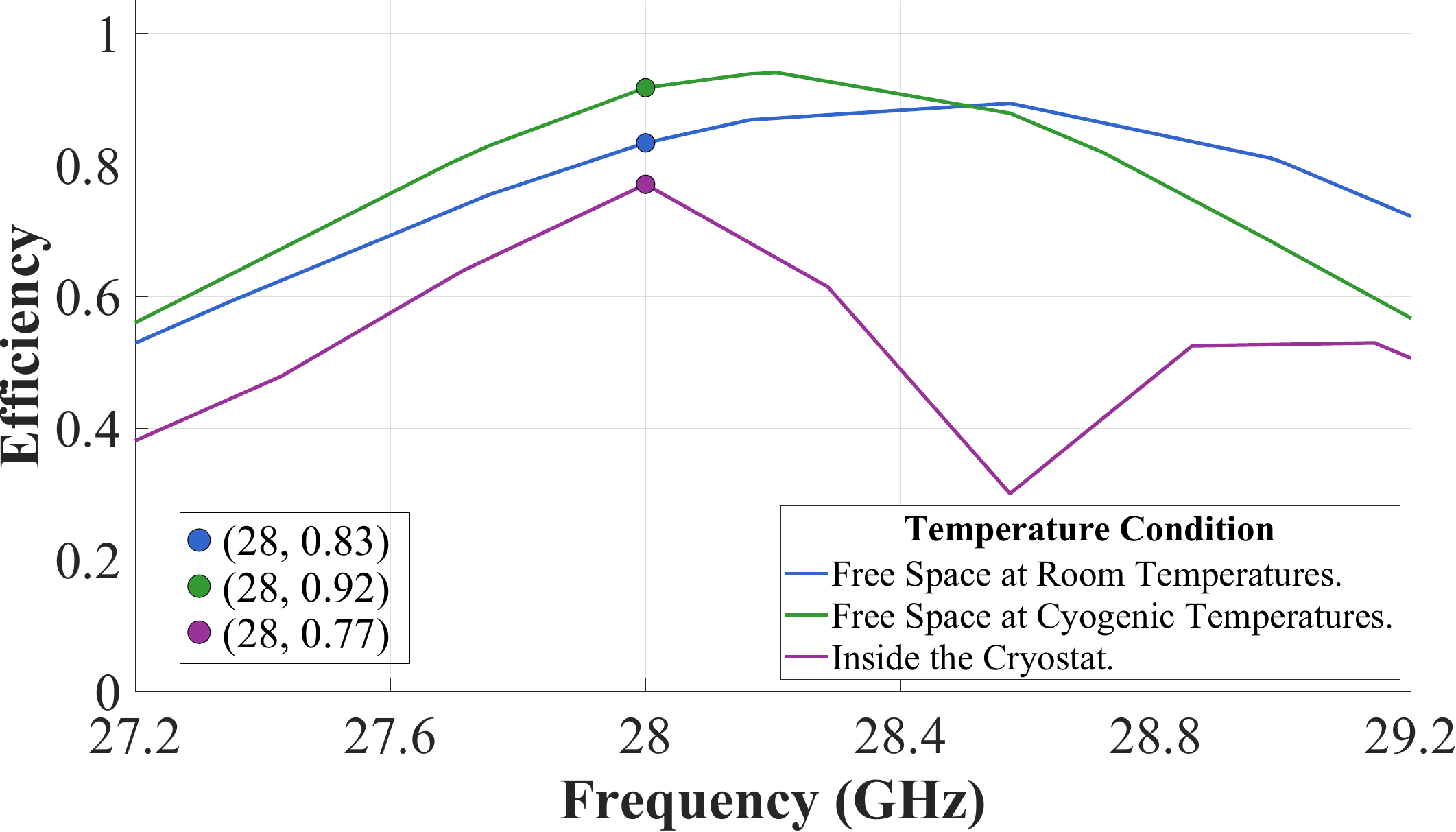}
\caption{}
\label{fig:s11cryostat}
\end{subfigure}
\vspace{-0.1cm}
\caption{Performance of the on-chip differential dipole antenna at 28 GHz. Comparison across free space under room and cryogenic temperatures, and inside the cryostat, showing: (a) Reflection coefficient (dB) (b) Total efficiency }
\label{fig:snrcap}
\vspace{-0.2cm}
\end{figure*}

Changes in material properties due to different temperature conditions resulted in reduced energy dissipation and improved antenna conductivity in cryogenic environment. Fig. \ref{fig:roomvscryo} shows the $S_{11}$ parameter of the on-chip antenna at room temperature, which has a value of -21.81 dB at 28 GHz. The variations of material properties collectively boost the efficiency of the antenna, raising it from 83\% to 92\% when the temperature decreases.

\begin{figure*}[!t]
\centering
\begin{subfigure}[t]{0.27\textwidth} 
\includegraphics[width=1\textwidth]{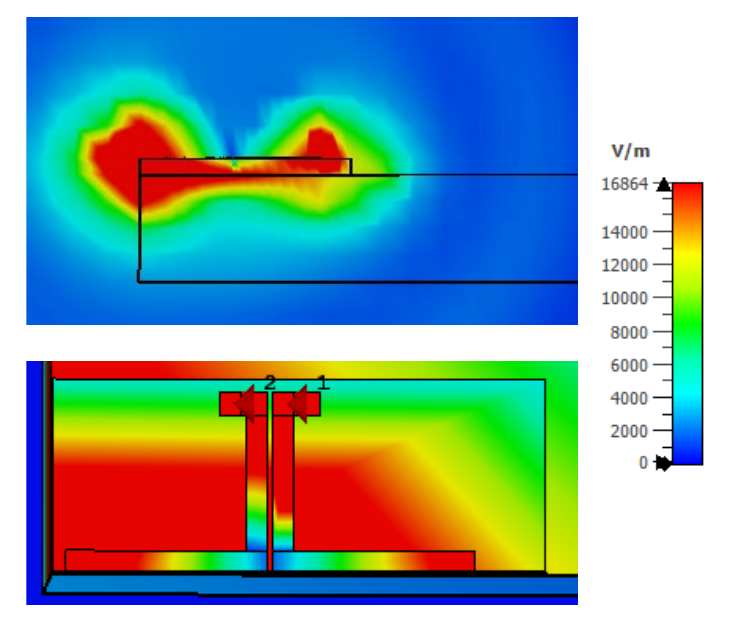}
\caption{}
\label{fig:efield}
\end{subfigure}
\begin{subfigure}[t]{0.4\textwidth} 
\includegraphics[width=1\textwidth]{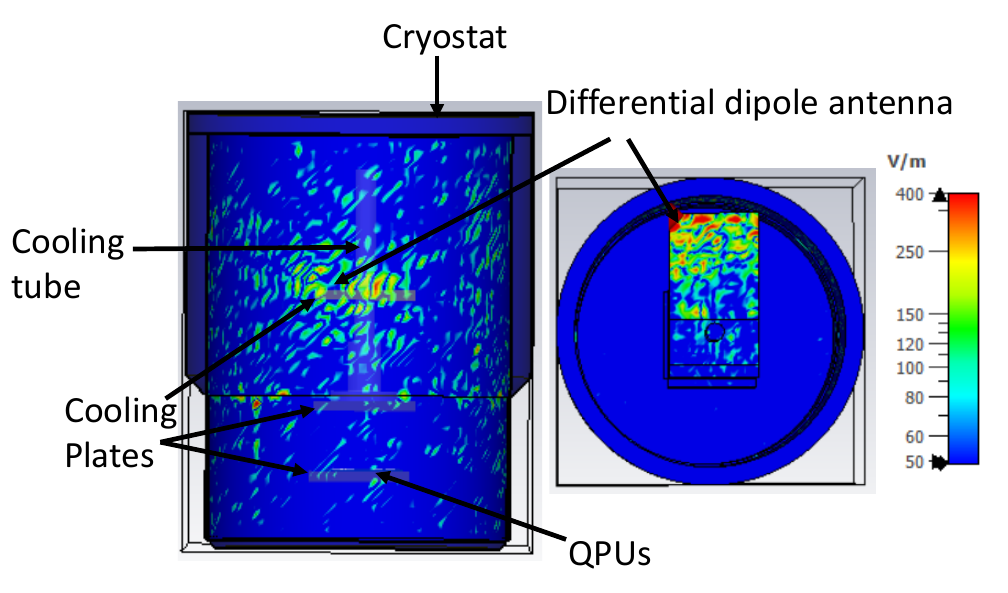}
\caption{}
\label{emcryo}
\end{subfigure}
\vspace{-0.1cm}
\caption{Performance of the differential dipole antenna in 28GHz (a) E-field distribution of the on-chip differential dipole inside the cryostat at 28 GHz (b) Spatial field distribution inside the cryostat, where the height=70 cm and the diameter=30 cm. }
\label{fig:snrcap}
\vspace{-0.2cm}
\end{figure*}

\subsection{Cryo-CMOS antenna within the Cryostat Scenario} 

To ensure proper impedance matching and efficient power transfer, the free-space dipole optimal length ($L_{opt}$) of 2.8 mm was replaced with 3.06 mm. In addition, the length of the SiO\textsubscript{\textrm{2}} and Si layers ($L_{layers}$), as well as the thickness of the substrate, were explored. An optimal resonance was obtained with $L_{layers}$ of 3.72 mm and Si thickness of 0.3 mm. As depicted in Fig. \ref{fig:roomvscryo}, the on-chip dipole resonates at around 28 GHz with an approximate value of -21 dB. The efficiency and realized gain of the antenna were also assessed inside the cryostat. Fig. \ref{fig:s11cryostat} shows an efficiency of 77\%, which reflects the effects of the cryostat environment on the antenna resonance. In addition, a realized gain of 8.25 dBi was obtained at 28~GHz. However, thermal asymmetries near the differential feed lines can trigger a common-mode conversion, which have to be carefully addressed in cryogenic conditions.

The EM field distribution of the on-chip antenna was also simulated. Fig. \ref{fig:efield} depicts the signal propagating from the GSSG pads to the dipole arms, the coupling and signal propagation around the dipoles, where the differential feeding introduces a 180° phase shift between them. Additionally, EM field distribution inside the cryostat model is observed in Fig. \ref{emcryo}. It is shown that the electric fields are oscillating around the enclosed chamber with multi-path signal propagation. 
However, as the QPUs are placed one temperature level below, the impact of radiated power is reduced.

\section{Conclusion}
\label{sec:conclusion}

In this paper, we propose and evaluate an on-chip differential dipole antenna designed for operation at 28 GHz within a quantum computer structure. By leveraging cryo-CMOS technology and material properties adapted for cryogenic environments, the dipole exhibits enhanced performance compared to room temperature, achieving reflection coefficients of -28.85 dB in free space with and -21 dB inside the cryostat corresponding to total efficiencies of 92\% and 77\% respectively. The simulations further demonstrate the impact of substrate thickness, layer dimensions on the antenna performance while initially assessing the EM distribution inside a realistic cryostat computer.

\bibliographystyle{IEEEtran}
\bibliography{IEEEabrv,bib}

\end{document}